\documentclass[11pt,epsfig,epsf,amsfonts,amsmath,enumerate,amssymb,amsthm,graphicx]{article}
\usepackage{amsthm}
\usepackage{amssymb}
\usepackage{epic}
\usepackage{eepic}
\usepackage{graphics}
\usepackage{epsfig}
\usepackage{psfrag}

\topmargin by -\headsep \textheight 8.9in \oddsidemargin 0pt
\evensidemargin \oddsidemargin \marginparwidth 0.5in \textwidth
6.5in

\def\proof{\par\vskip 3pt\noindent\hbox{\bf Proof} :\quad}
\newtheorem*{thm*}{Theorem}
\newtheorem{thm}{Theorem}[section]

\newtheorem{lem}{Lemma}[section]
\newtheorem*{lem*}{Lemma}

\newtheorem{cor}{Corollary} [section]

\def\qed{\hfill\raisebox{3pt}{\fbox{\rule{0mm}{1mm}\hspace*{1mm}\rule{0mm}{1mm}}\,}
\vspace{8pt}}
\def\cadre{$$\vcenter\bgroup\advance\hsize by -8em\noindent
             \refstepcounter{equation}\ignorespaces}
\makeatletter
\def\endcadre{\egroup\eqno(\theequation)$$\global\@ignoretrue}
\makeatother

\def\Qset{\hbox{\hbox{Q\hskip-0.525em\lower-0.097ex
\hbox{\vrule height1.47ex width 0.07em}}\hskip0.50em}}

\newcounter{petit}

  \def\Pscr{{\cal P}}

\def\Qset{\hbox{\hbox{Q\hskip-0.525em\lower-0.097ex
\hbox{\vrule height1.47ex width 0.07em}}\hskip0.50em}}

\begin{document}
		\vskip -0.93cm
\title{Ear-Slicing for Matchings in  Hypergraphs}
\author{Andr\'as Seb\H{o}  \\CNRS, Laboratoire G-SCOP,
	Univ. Grenoble Alpes}

\maketitle
\begin{abstract} We study when a given edge of a factor-critical graph is contained in a matching avoiding exactly one, pregiven vertex of the graph.   We then apply the results to always partition the vertex-set of  a $3$-regular, $3$-uniform hypergraph into at most one triangle (hyperedge  of size $3$) and edges (subsets of size $2$ of hyperedges), corresponding to the intuition, and providing new insight to  triangle and edge packings of  Cornu\'ejols' and Pulleyblank's.  The existence of such a packing can be considered to be a hypergraph variant of Petersen's theorem on perfect matchings, and  leads to a simple proof for a sharpening of Lu's theorem  on antifactors of graphs. 
\end{abstract}

\section{Introduction }\label{sec:intro}

 Given a hypergraph $H=(V,E)$,  $(E\subseteq\Pscr(V),$ where $\Pscr(V)$ is the power-set of $V)$  we will call the elements of $V$ {\em vertices}, and those of $E$ {\em hyperedges}, $n:=|V|$.  We say that a  hypergraph is {\em $k$-uniform}  if all of its hyperedges have $k$ elements, and it is  {\em $k$-regular} if all of its vertices are contained in $k$ hyperedges. Hyperedges may be present with multiplicicities, for instance the hypergraph  $H=(V,E)$ with $V=\{a,b,c\}$ consisting of the hyperedge $\{a,b,c\}$ with multiplicity $3$ is a $3$-uniform, $3$-regular hypergraph.
 The  {\em  hereditary closure} of the hypergraph $H=(V,E)$ is $H^h=(V,E^h)$ where $E^h:=\{X\subseteq e: e\in E\}$, and $H$ is {\em hereditary}, if $H^h=H$. For the new edges of the hereditary closure we do not need to define multiplicities we will consider them all to be one.
  

 Hyperedges of cardinality $1$ will be called {\em singletons}, those of cardinality $2$ and $3$ are called {\em edges} and {\em triangles} respectively.   Deleting a vertex $v$ of the hypergraph $H$ results in the hypergraph $H-v=(V\setminus \{v\}, \{e\in E, v\notin E\} )$. For hereditary hypergraphs this is the same as deleting $v$ from all hyperedges. The degree $d_H(v)$ of $v\in V$ in $H$ is the number of hyperedges containing $v$.  

Given a    hypergraph  $H=(V,E)$, denote by $E_2$ the set of edges (of size two) in $E^h$, $H_2:=(V,E_2)$. We do not need parallel edges in  $H_2$, we suppose $H_2$ is a graph without parallel edges or loops.   The (connected) {\em components} of $H$ are defined as those of $H_2$. These form a partition of $V$, and correspond to the usual hypergraph components: $H$ is {\em connected} if $H_2$ is connected. Abusing terminology, the vertex-set of a component is also  called component. We define a {\em graph} as a hypergraph $H$ with $H=H_2$, that is, a $2$-uniform hypergraph without loops or parallel edges.

A {\em matching} in a graph is a set of pairwise vertex-disjoint edges. A matching is {\em perfect} if it partitions the vertex-set of the graph. 
A graph $G=(V,E)$ is called {\em factor-critical} if $G-v$ has a perfect matching (also called a $1$-factor) for all $v\in V$. 

In this note we prove two lemmas, possibly interesting for their own sake, on when a given edge of a factor-critical graph is contained in a matching avoiding exactly one, pregiven vertex of the graph, leading to a result on $3$-uniform hypergraphs (Section~\ref{sec:factorcrit}). 
 We then prove that a $k$-regular and $k$-uniform hypergraph is perfectly matchable in some sense (a generalization of Petersen's theorem \cite{Petersen}  on $2$-uniform hypergraphs),  sharpening a result of Lu's  \cite{Lu}    (Section~\ref{sec:main}). 

\section{Ears and Triangles}\label{sec:factorcrit} 
An {\em ear-decomposition} $P_0+\ldots + P_k$ consists of   a circuit
$P_0$, and paths $P_i$ $(i\in\{1,\ldots,k\})$  sharing  its (one or two)  {\em endpoints} with $V(P_0)\cup\cdots\cup V(P_{i-1})$;
$P_1,\ldots,P_k$ are called {\em ears}.
An ear is called {\em trivial}, if it consists of one edge. An ear is called {\em odd} if it has an odd number of edges.
Lov\'asz \cite{LL}, \cite{LPl}  proved that a graph is factor-critical if and only if it has an  ear-decomposition with all ears odd. 

For $v\in V$, denote by ear$(v)$ the index of the first ear when vertex $v$ occurs. (It may occur later only as an endpoint of an ear.)  Given an ear-decomposition, we call an edge $e=ab$ {\em odd}, if ear$(a)=~$ear$(b)$, and if $i\ge 1$ we also require that $a$ is joined to an endpoint of $P_i$ by an odd subpath of $P_i$ not containing $b$, and that the same holds interchanging the role of $a$ and $b$. We will call an odd ear-decomposition {\em maximal} if for every odd edge $ab$, $i:=~$ear$(a)=~$ear$(b)$, we have $ab\in P_i$.  

Clearly, {\em there exists a maximal odd ear-decomposition, since while there are odd edges $e\notin P_i$ with endpoints on $P_i$ $(i\in\{0,\ldots,k\})$, we can obviously replace the ears $P_i$ and $e$, where $\{e\}$ is necessarily a trivial ear, with two odd ears.} In particular, an odd ear-decomposition with a maximum number of nontrivial ears (equivalently, with a minimum number of trivial ears) is maximal.  

We need the following  lemmas that may also have some self-interest and other applications:  for  $v\in V$, $e\in E$, it provides a sufficient condition for $G-v$ to have a perfect matching containing $e$. 

\begin{lem}\label{lem:oddedge}
	Let  $G=(V,E)$ be a factor-critical graph given with an odd ear-decomposition, and let $e=ab\in E$ be  an odd edge in a nontrivial ear.  For any vertex $v\in V$ with 
	{\rm ear}$(v)<~${\rm ear}$(a)=~${\rm ear}$(b)$,  there exists a perfect matching of $G-v$  containing $e$. 
\end{lem}
\proof Let  $G=P_0+\ldots+ P_k$ be the ear-decomposition. We can suppose without loss of generality (since by the easy direction of Lov\'asz's theorem \cite{LL}, a graph having an odd ear-decomposition is factor-critical) that  $e$ is on the last ear $P_k$.   Since $v$ is in  $G'=P_0\cup\ldots\cup P_{k-1}$, and $G'$ is factor-critical (again by the easy direction of Lov\'asz's theorem),  $G'-v$ has a perfect matching $M'$. Adding the odd edges of $P_k$ to $M'$, we get the matching of the assertion. \qed


Cornu\'ejols, Hartvigsen and Pulleyblank \cite{CP}, \cite{CHP} (see also \cite{LPl}) need to check when a factor-critical graph is partitionable into triangles and edges, and for this they try out all triangles. The following lemma improves this for the unions of triangles by showing that they {\em always} have such a partition:
 
 \begin{lem}\label{lem:onetriangle}
	If $H=(V,E)$ is a $3$-uniform hypergraph and $H_2$ is factor-critical, then $V$ has a partition into one triangle $\{a,b,c\}\in E$ and a perfect matching in $H_2-\{a,b,c\}$. 
\end{lem}
\proof Consider a maximal odd ear-decomposition, let $e=ab\in E_2$ be an odd edge on its last nontrivial ear $P_k$ and let $\{a,b,v\}\in E$ be a triangle containing $e$.  If ear$(v)<k$, we are done by Lemma~\ref{lem:oddedge}: a matching $M$, $e\in M$ of $G-v$ and the triangle $\{a,b,v\}$ do partition  $V$.  Suppose now 
ear$(v)=k$.

If $k=0$ choose a vertex on $P_0$ at odd distance from both $a$ and $b$, and let both $u_1$ and $u_2$ denote this same vertex. The following proof holds then for both $k=0$ or $k\ge 1$. 

   We can suppose without loss of generality that the endpoint   $u_1$  of the ear $P_k$, $v$, $a$, $b$ and the other endpoint $u_2$ of $P_k$ (possibly $u_1=u_2$) follow one another in this order on the ear. The path between $u_1$ and $v$ is even, because if it were odd, the edge $vb$ would be odd - the path between $b$ and $u_2$ being odd by the assumption that the edge $ab$ is odd -, contradicting the maximality of the ear-decomposition. But then a perfect matching of $(P_0+\ldots+P_{k-1}) - u_1$ and every second edge of the subpath of $P_k$ between $u_1$ and $v$, covering $u_1$ but not covering $v$, and the odd edges of the rest of $P_k$ including $e$, form a perfect matching in $H_2 - v$ containing $e$. Replacing $e$ in this perfect matching by $\{a,b,v\}\in E$  finishes the proof.
\qed 

\section{Regular Hypergraphs}\label{sec:main}
 \begin{thm}\label{thm:main}
	If $H=(V,E)$ is a $3$-uniform, $3$-regular  hypergraph, then $H_2$ has either a perfect matching (if $|V|$ is even), or it is factor-critical (if $|V|$ is odd) and in the latter case $V$ can be partitioned into one triangle $\{a,b,c\}\in E$ and a perfect matching of $H_2-\{a,b,c\}$. 
\end{thm}
\proof If $H_2$ has a perfect matching we have nothing to prove.  Suppose it has not.

\smallskip
\noindent {\bf Claim.} $H_2$ is factor-critical.

We prove more: {\em for $X\subseteq V$, $X\ne\emptyset$, the number $k$ of components of $G-X$ satisfies $k\le|X|$.} (Then by Tutte's theorem \cite{Tutte}, see also \cite{LPl}, $G-v$ has a perfect matching for all $v\in V$.)
For each component $C$, by $3$-regularity,    $\sum_{v\in C}d_H(v)=3|C|$. 

In this sum, divisible by $3$, every hyperedge is counted as many times as it has vertices in $C$. Since $G$ is connected, and $C\ne V$, we have that the sum of $|e|$ for all hyperedges that meet $C$,  itself divisible by $3$ by $3$-uniformity,  is  strictly larger than $\sum_{v\in C}d_H(v)$. Therefore, the sum of $|e\setminus C|$ for these edges is nonzero and also divisible by $3$, so it is at least $3$. 
The vertices not in $C$ of the edges that meet $C$ are in $X$, hence $e\setminus C=e\cap X$, so summing $|e\cap X|$ for all edges, the sum is at least $3k$:  
\[3k \le \sum_{e\in E}|e\cap X|\le \sum_{x\in X} d_H(x) = 3|X|,\] 
so $k\le |X|$, finishing the proof of the claim. Now Lemma~\ref{lem:onetriangle} can be readily applied. 
\qed

The intuition that at most one triangle may be enough is highly influenced by Cornu\'ejols, Hartvigsen and Pulleyblank's work \cite{CP}, \cite{CHP}, even if these are not explicitly used. The heart of the proof is encoded in the two lemmas that show: we can either increase the number of nontrivial ears or find the wanted partition, and for this, $3$-uniformity is not needed. The proof is clearly algorithmic, providing a low degree polynomial algorithm. 

Finally, we prove a sharpening of Lu's theorem \cite{Lu}, which considered a question in \cite{LPl}. A simple proof of Lu's theorem has been the initial target of this work. 

\begin{cor}\label{cor:Lu} Let $G$ be a $k$-regular bipartite graph with bipartition $\{A,B\}$ and $k\in\mathbb{Z}, k\ge 3$. Then $G$ has a subgraph with all degrees of vertices in $B$ equal to $1$, all degrees of vertices in $A$ equal to $2$ or $0$, except possibly at most one vertex of $A$ which is of degree $3$.   
\end{cor}

\proof Delete $k-3$ pairwise disjoint perfect matchings one by one (they are well-known to exist in bipartite regular graphs~ \cite{LPl} by Hall's theorem, actually a $k$-edge-coloring also exists by K\H{o}nig's edge-coloring theorem). Define then the hypergraph $H=(V,E)$ with $V:=B$, and $E$ to have one hyperedge for each $a\in A$ consisting of  the set of neighbors of $a$. Since there are no loops or parallel edges in $G$ (see Section~\ref{sec:intro}), the defined hypergraph $H$ is $3$-uniform and $3$-regular.   

Apply now Theorem~\ref{thm:main}. 
\qed

As the proof   shows, the essential case is $k=3$, when the theorem can be considered to be a generalization of Petersen's theorem \cite{Petersen} about perfect matchings in  graphs. Let us also state the reformulation to hypergraphs by the inverse of the correspondence in the proof: 

\begin{cor}\label{cor:k} If $H=(V,E)$ is a $k$-uniform, $k$-regular  hypergraph, $k\in\mathbb{Z}, k\ge 3$,  then $V$ can be partitioned into hyperedges of $H^h$ of size  $2$ and at most one hyperedge of size $3$. 
	\end{cor}


\noindent
{\bf Acknowledgment:} {Many thanks to Zolt\'an Szigeti and Louis Esperet for precious suggestions!}

\end{document}